\documentclass[twocolumn,amsmath,amssymb,floatfix,prl]{revtex4-1}
\usepackage{amsmath}
\usepackage{amssymb}
\usepackage{graphicx}
\usepackage{color}

\def\Rb87{^{87}\rm{Rb}}				
\def\Rb85{^{85}\rm{Rb}}				
\def\Cs133{^{133}\rm{Cs}}				
\def\Li6{^{6}\rm{Li}}					
\def\Li7{^{7}\rm{Li}}					

\begin{document}
\title{Geometric Scaling of Efimov States in a $^{6}$Li-$^{133}$Cs Mixture\\}

\author{Shih-Kuang~Tung}
\author{Karina~Jim\'{e}nez-Garc\'{\i}a}
\author{Jacob Johansen}
\author{Colin V. Parker}
\author{Cheng Chin}
\email{cchin@uchicago.edu}
\affiliation{The James Franck Institute, The Enrico Fermi Institute, and Department of Physics, The University of Chicago,
Chicago, Illinois 60637, USA}

\date{\today}

\begin{abstract}
In few-body physics, Efimov states are an infinite series of three-body bound states that obey universal discrete scaling symmetry
when pairwise interactions are resonantly enhanced. Despite abundant reports of Efimov states in recent cold atom experiments,
direct observation of the discrete scaling symmetry remains an elusive goal. Here we report the observation of three consecutive
Efimov resonances in a heteronuclear Li-Cs mixture near a broad interspecies Feshbach resonance. The positions of the resonances
closely follow a geometric series $1$, $\lambda$, $\lambda^2$. The observed scaling constant $\lambda_{\rm exp} = 4.9(4)$ is in
good agreement with the predicted value of 4.88.
\end{abstract}

\maketitle

The emergence of scaling symmetry in physical phenomena suggests a universal description that is insensitive to microscopic
details. Well-known examples are critical phenomena, which are universal and invariant under continuous scaling
transformations~\cite{Stanley1987}. Equally intriguing are systems with discrete scaling symmetry, which are invariant
under scaling transformations with a specific scaling constant~\cite{Sornett1998}; a classic example is the self-similar
growth of crystals, as in snowflakes. Surprisingly, such discrete scaling symmetry also manifests in the infinite series
of three-body bound states that Vitaly Efimov predicted in 1970~\cite{Efimov1970}.

\begin{figure}
 \begin{center}
 \includegraphics[width=3.4in]{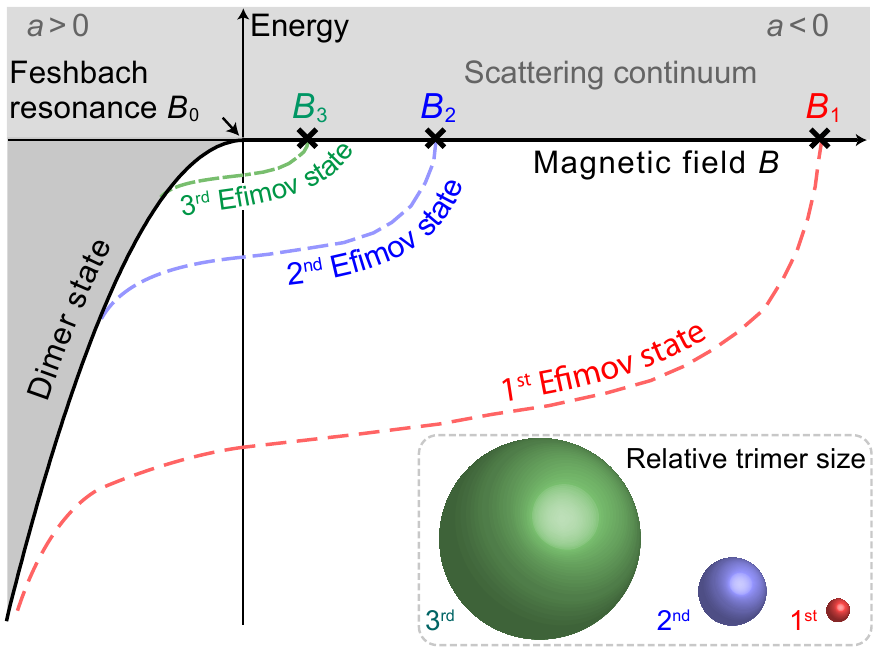}\\
 \end{center}
 \vspace{-10pt}
 \caption{{\bf Discrete scaling symmetry of Efimov states}. A series of Efimov states (dashed curves) exists near a Feshbach resonance located at $B_0$. Away from the resonance, on the side with scattering length $a<0$, they merge into the three-body scattering continuum (crosses). Physical observables in the Efimov scenario --- including the molecular size $R_n$, the binding energies $E_n$, and the Efimov resonance positions at the scattering length $a_n$ (associated with the magnetic field $B_n$) --- show discrete scaling symmetry. The discrete scaling law is graphically represented by the location of the resonances on the $B$-axis and the size of the spheres, respectively.}
 \label{fig:Figure1}
\end{figure}

In the Efimov scenario, while pairs of particles with short-range resonant interactions cannot be bound, there exists an infinite series of three-particle bound states. These bound states have universal properties that are insensitive to the details of the molecular potential and display discrete scaling symmetry; the size $R_n$ and binding energy $E_n$ of the Efimov state with the $n$th lowest energy scale geometrically as $R_{n} = \lambda R_{n-1}$ and $E_{n}=\lambda^{-2} E_{n-1}$, where $\lambda$ is the scaling constant. An alternative picture to understand discrete scaling symmetry is based on renormalization group limit cycles~\cite{Braaten2006}. Away from the two-body scattering resonance, Efimov states couple to the scattering continuum and induce a series of three-body scattering resonances at scattering lengths $a_{-}^{(n)}<0$, which also follow the scaling law $a_{-}^{(n)}=\lambda a_{-}^{(n-1)}$~\cite{Esry1999} (Fig.~\ref{fig:Figure1}).

Ultracold atom systems are ideal to test Efimov scaling symmetry given that their interatomic interactions can be tuned over several orders of magnitude using Feshbach resonances~\cite{Chin2010}. The first evidence of an Efimov state was reported in ultracold Cs atoms~\cite{Kraemer2006}; subsequent observations of Efimov resonances~\cite{note1} in homonuclear systems were also reported in $^7 \rm Li$~\cite{Pollack2009, Gross2009}, $^{39}\rm K$~\cite{Zaccanti2009}, $^{85} \rm Rb$~\cite{Wild2012}, $^{133} \rm Cs$~\cite{Berninger2011, Huang2014}, and $^6 \rm Li$~\cite{Ottenstein2008, Huckans2009}. Despite these numerous observations, experimental confirmation of discrete scaling symmetry remains a challenging goal.

The confirmation of universal discrete scaling symmetry requires the observation of multiple Efimov resonances. With two consecutive Efimov resonances, the scaling symmetry can be tested through a comparison between the ratio of the resonance positions and theory; with three or more resonances one can perform a model-independent test. The observation of multiple resonances is experimentally challenging because higher order Efimov resonances diminish when the scattering rate is unitarity limited~\cite{Kraemer2006, Rem2013, Fletcher2013}. This challenge is acute in homonuclear systems with large scaling constant $\lambda\approx 22.7$~\cite{Efimov1970}, such that the detection of an additional Efimov state demands a reduction of temperature by a factor of $\lambda^2\approx515$. Features from excited Efimov states in homonuclear systems have been observed in $^{6} \rm Li$ \cite{Williams2009}, $^{39}\rm K$~\cite{Zaccanti2009},  $^7 \rm Li$~\cite{Dyke2013}, and $^{133}$Cs~\cite{Huang2014}. These results are consistent with the scaling prediction, but do not provide an independent test of the scaling symmetry.

Heteronuclear systems consisting of one light atom resonantly interacting with two heavy atoms can have a scaling constant significantly lower than 22.7~\cite{DIncao2006, Braaten2006, Helfrich2010, Wang2012}; however, experiments in heteronuclear systems are considerably more challenging than those in homonuclear systems. Before our work, observations of Efimov resonances in heteronuclear systems were reported in K-Rb mixtures~\cite{Barontini2009,Bloom2013}. Recently, in a Li-Cs mixture~\cite{Pires2014}, two Efimov resonances are found in the measurement of three-body loss coefficients, and the number loss data hint at the existence of a third Efimov resonance.

Here we report the observation of discrete scaling symmetry of Efimov states in a Fermi-Bose mixture of $^6 \rm Li$ and $^{133} \rm Cs$. Taking advantage of the large mass ratio between Li and Cs atoms, with a predicted scaling constant $\lambda =4.88$~\cite{DIncao2006, Wang2012}, we identify three consecutive Efimov resonances near a wide, isolated $s$-wave interspecies Feshbach resonance~\cite{Tung2013}. From the measured locations of the resonances, we provide a model-independent proof of the geometric scaling symmetry and determine a scaling constant $\lambda_{\rm exp} = 4.9(4)$.

Our experiment is based on a mixture of $^6$Li and $^{133}$Cs atoms near quantum degeneracy in an optical dipole trap. In our experiment, both species are prepared in their lowest states, $\left| F = 1/2, m_F=1/2 \right>$ for Li and $\left|F=3,m_F=3\right>$ for Cs, where $F$ is the total angular momentum and $m_F$ is its projection. We prepare mixtures with up to $N_{\rm Li}=3.4 \times 10^4$ Li atoms, and $N_{\rm Cs} = 5.2\times 10^4$ Cs atoms at temperatures in the range $190 \;{\rm nK} <T<800\;{\rm nK}$~\cite{Methods2014}. Efimov resonances manifest themselves as enhanced three-body recombination losses. In such collisions three atoms resonantly couple to an Efimov state and then decay into a deeply bound molecule and a free atom; the released binding energy allows them to escape the trap. We measure the Li and Cs atom numbers from which we infer atom loss and identify the Efimov resonances.

The mixture of $^6$Li and $^{133}$Cs has two primary inelastic collision pathways: three-body recombination of Cs-Cs-Cs and Li-Cs-Cs. At low temperatures, Li-Li-Cs as well as Li-Li-Li collisions are strongly suppressed by Fermi statistics. We investigate Efimov resonances near the broad Li-Cs Feshbach resonance located at 842.75~G with a width of 61.6~G and a strength parameter $s_{\rm res}$ of $\sim$ 0.7~\cite{Tung2013}. The Efimov resonances reported in this work are away from $p$-wave Feshbach resonances~\cite{Repp2013}, as well as Cs Feshbach and Efimov resonances~\cite{Berninger2011,Berninger2013}.

\begin{figure}
 \begin{center}
 \includegraphics[width=3.4in]{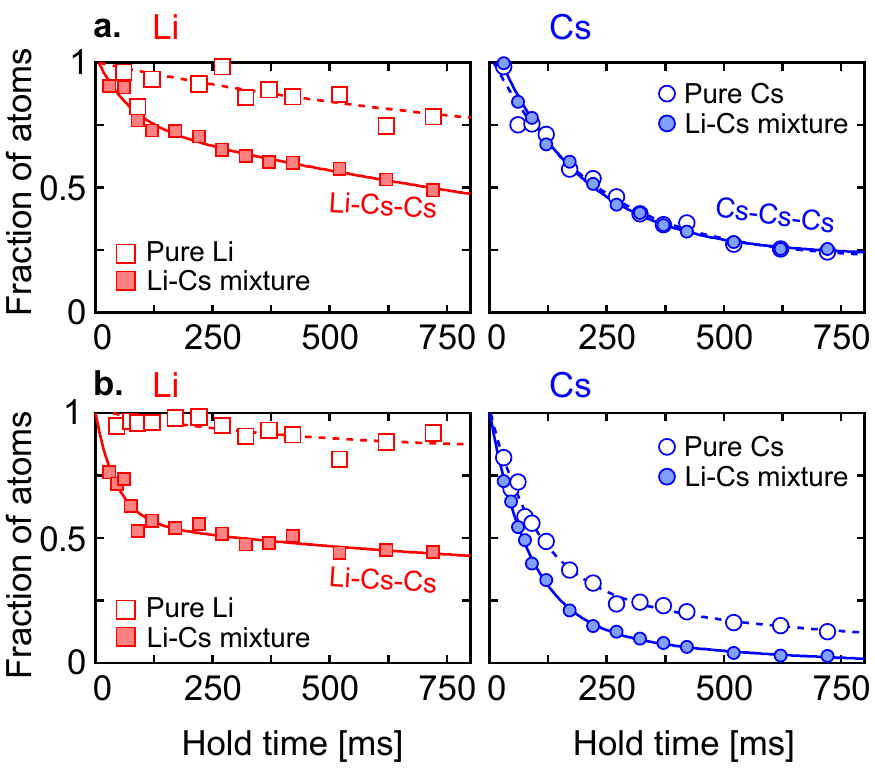}\\
 \end{center}
 \vspace{-10pt}
 \caption{{\bf Atom number decay of single-species and Li-Cs mixture samples}. {\bf a.} At $B=$848.0~G, $\sim$5~G away from the Li-Cs Feshbach resonance ($a_{\rm LiCs}=-354~a_0 $, $a_{\rm CsCs}=-1240\;a_0$) Li loss increases significantly when Cs is introduced (left panel). Cs loss is dominated by Cs-Cs-Cs recombination (right panel). Sample temperature is $T= 390$~nK. {\bf b.} Near the Li-Cs Feshbach resonance $B=842.7$~G ($a_{\rm CsCs}=-1570~a_0$), enhanced atom loss is evident in both the Li and Cs atom number evolution when both species are present. Sample temperature is $T= 340$~nK. Data in {\bf a.} and {\bf b.} are scaled to the initial atom numbers, $N_{\rm Li} = 2\sim3\times10^4$ and $N_{\rm Cs} = 4\sim5\times10^4$, obtained from double exponential fits (continuous and dotted lines), which also serve as guides to the eye.}
 \label{fig:Figure2}
\end{figure}

Around the magnetic field region probed in this work, the Cs-Cs scattering length is large and negative, and Cs-Cs-Cs recombination is the major competing loss process, imposing a limitation on the lifetime of Cs. Away from the Li-Cs Feshbach resonance, Cs decay is dominated by Cs-Cs-Cs recombination collisions [Fig.~\ref{fig:Figure2}(a)]; on the other hand, Li decays much faster in the presence of Cs, indicating the dominance of interspecies collisional loss. Near the Feshbach resonance [Fig.~\ref{fig:Figure2}(b)] both decays of Li and Cs are enhanced by interspecies collisions.

\begin{figure*}
 \begin{center}
 \includegraphics[width=6.5in]{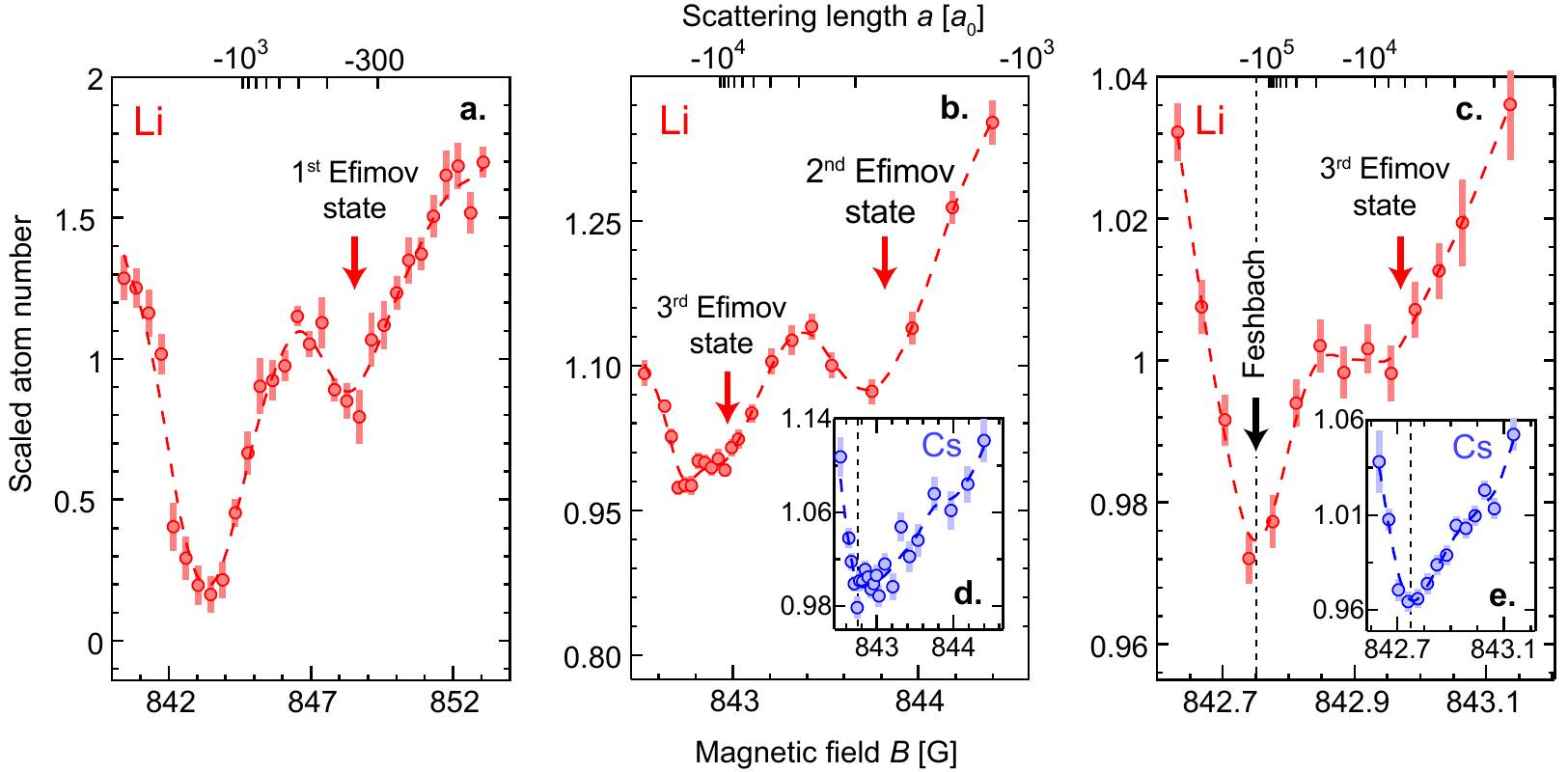}\\
 \end{center}
 \vspace{-10pt}
 \caption{{\bf Observation of three Li-Cs-Cs Efimov resonances.} {\bf a.} Scaled Li number versus magnetic field showing the first Li-Cs-Cs Efimov resonance, from the average of 13 individual scans. Here ${N_{\rm Li} = 1.3\times 10^4}$ and ${N_{\rm Cs} = 2.7\times 10^4}$ with typical temperature $T=800$~nK and hold time 225 ms. {\bf b.} Scaled Li and Cs (inset, {\bf d}) numbers near the second and third Li-Cs-Cs Efimov resonances, from the average of 68 scans with  typical temperature $T= 360$~nK and hold time 115 ms. The mean atom numbers are $N_{\rm Li} = 1.4\times 10^4$ and $N_{\rm Cs} = 2.1\times 10^4$. {\bf c.} Scaled Li and Cs (inset, {\bf e}) numbers  close to the third Li-Cs-Cs Efimov resonance and the Li-Cs Feshbach resonance, from the average of 327 scans with typical temperature $T= 270$~nK and hold time 115 ms. The mean atom numbers are $N_{\rm Li} = 9.1\times 10^3$ and $N_{\rm Cs} = 1.4 \times 10^4$. The scaled atom numbers come from the average of the individual scans divided by their respective mean values. The vertical dashed lines indicate the Feshbach resonance and arrows indicate the Efimov resonances. The dashed curves correspond to an interpolation of the data and serve as a guide to the eye.}
\label{fig:Figure3}
\end{figure*}

Our measurements of atom loss and observation of Efimov resonances are summarized in Fig.~\ref{fig:Figure3}. To eliminate the long-term drift in atom number, we scale the atom number so that it averages to unity over a fixed magnetic field range.  Each panel shows resonant loss features in the scaled atom number. The main loss feature in both Li and Cs scans is associated with a broad Li-Cs Feshbach resonance \cite{Tung2013, Repp2013}. Loss features associated with excited Efimov resonances on the negative scattering length side of the Feshbach resonance are only evident in low temperature scans, and indicated by arrows in Fig.~\ref{fig:Figure3}. Efimov features are weaker in Cs data due to fast competing Cs-Cs-Cs recombination processes.

We determine the position of each Efimov resonance by using both Lorentzian and Gaussian fits with a linear background. Results from different fit functions and fit ranges are analyzed and combined to determine the final resonance positions and uncertainties. Further details on the fit and the determination of the resonance positions and uncertainties are given in Ref.~\cite{Methods2014}. We determine the positions of the three  Efimov resonances to be $B_1$ = 848.55(12)$_{\rm stat}$(3)$_{\rm sys}$~G, and $B_2$ = 843.82(4)$_{\rm stat}$(3)$_{\rm sys}$~G, and $B_3$ = 842.97(3)$_{\rm stat}$(3)$_{\rm sys}$~G, where $(\;)_{\rm stat}$ denotes the statistical uncertainty and the systematic uncertainty of 30~mG arises from the daily magnetic field drift.

A precise determination of the Feshbach resonance position is crucial to check the scaling symmetry. Two independent methods are developed here. First, we observe that the strongest dip (Fig.~\ref{fig:Figure3}) is ubiquitous in all measurements, even at high temperatures where Efimov features are indiscernible. This indicates that the strongest dip is associated with the Feshbach resonance. Fits to the lowest temperature data [Fig.\ref{fig:Figure3}(c)] locate the Feshbach resonance at $B_0=$842.75(1)$_{\rm stat}$(3)$_{\rm sys}$~G. 

\begin{figure}
 \begin{center}
 \includegraphics[width=3.4in]{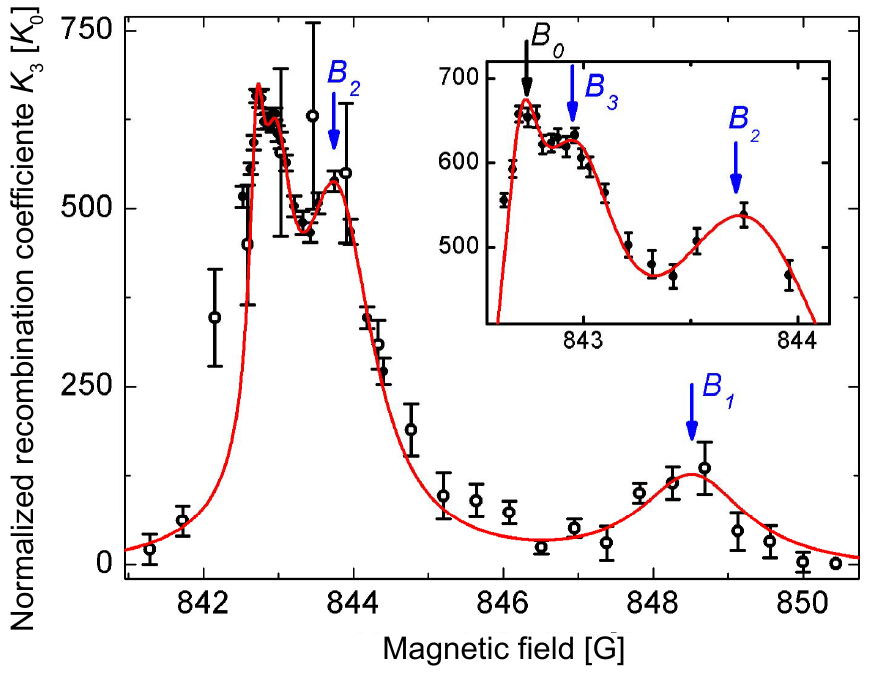}\\
 \end{center}
 \vspace{-10pt}
 \caption{{\bf Feshbach and Efimov resonance structure in recombination coefficient}. We extract and normalize the recombination coefficient $K_3$ from the atom number measurement based on a rate equation model~\cite{Methods2014}. Using the data in Fig.~3(a) ($T=800$~nK, open circles) and Fig.~3(b)~($T=360$~nK, solid circles), we show that the extracted $K_3$ displays four peaks. The three peaks at magnetic fields $B_1, B_2$ and $B_3$ are associated with Efimov resonances, and the global maximum at the lower field $B_0$ represents the Feshbach resonance. $K_0=10^{-25} {\rm cm}^6/{\rm s}$ is the loss coefficient we obtain at $\sim$852~G. The inset shows the zoom-in view of the resonance structure from the lower temperature measurement, which offers higher resolution to the higher order Efimov resonances. The solid line represents a four-Lorentzian fit to the data, and serves as a guide to the eye.}
 \label{fig:Figure4}
\end{figure}

We convert our atom loss measurement into a spectrum of the recombination loss coefficient, see Fig.~\ref{fig:Figure4}, based on a rate equation model~\cite{Methods2014}. The spectrum shows clearly three Efimov resonance features and can be compared with theoretical calculation. In addition, after comparing the extracted $K_3$ with a model that captures the steep rise of $K_3$ for $a>0$~\cite{Methods2014}, we find the best agreement between the experiment and the model when $B_0$ = 842.75 $(1)_{\rm stat}$(3)$_{\rm sys}$~G. The results from both our methods to determine $B_0$ agree with each other.

The separations between the Efimov resonances and the Feshbach resonance ${\Delta B_n = B_n-B_0}$ are $\Delta B_1=5.80(12)$~G, $\Delta B_2=1.07(4)~$G, and $\Delta B_3= 0.22(3)~$G; the uncertainties include both statistical and systematic errors. Remarkably, they closely follow a geometric progression $\Delta B_1:\Delta B_2:\Delta B_3 \approx 1:1/5:1/5^2$ and provide direct evidence of the discrete scaling symmetry. (Note that $\Delta B_n\propto -1/a_{-}^{(n)}$ near the Feshbach resonance.) More precisely, using an updated scattering model for the Li-Cs Feshbach resonance~\cite{Methods2014}, we determine the Efimov resonances in scattering length to be $a_{-}^{(1)}  = -323(8)\;a_0$, $a_{-}^{(2)} = -1635(60) \;a_0$, and $a_{-}^{(3)}  = -7850(1100)\;a_0$, where $a_0$ is the Bohr radius. Two scaling constants are extracted: $\lambda_{21} =a_{-}^{(2)}/a_{-}^{(1)} =  5.1(2)$ and $\lambda_{32} = a_{-}^{(3)}/a_{-}^{(2)}  = 4.8(7)$, which mutually agree within uncertainty. The averaged scaling constant $\lambda_{\rm exp} = 4.9(4)$ is in good agreement with the predicted value $\lambda=4.88$ for LiCs$_2$ Efimov states~\cite{DIncao2006, Wang2012}.

Even though the observed scaling ratios are consistent with the predicted value, we would like to point out the practical factors that could contribute to differences between experiment and theory. The first Efimov resonance can be shifted by finite-range corrections given that it occurs at a scattering length near the van der Waals length of Cs-Cs ($r_{\rm Cs Cs}=101~a_0$) and Li-Cs ($r_{\rm Li Cs}=45~a_0$). The location of the Efimov resonance can also be shifted by finite temperature and finite trap size effects, which are stronger for excited Efimov states~\cite{DIncao2004, Nagerl2006}. Based on a closer inspection, our data do not detect a clear position shift of the third Efimov resonance due to finite temperature effect. Finite size effects are estimated to be negligible in our experiment~\cite{Methods2014}.

In conclusion, we observed three Efimov resonances in a Li-Cs mixture and extracted two scaling constants. From their mutual agreement and that with theoretical calculations, our results provide experimental evidence of discrete scaling symmetry of Efimov states. Based on our observations, an intriguing question is whether the discrete scaling symmetry can be tested in the original Efimov scenario of three identical bosons, which requires extremely low temperature. In addition, our result also hints at the persistence of discrete scaling symmetry when the scattering length diverges (unitary Bose gas)~\cite{Wang2011}, in 
contrast with the continuous scaling symmetry of a unitary Fermi gas~\cite{Ho2004}.

We thank Yujun Wang for valuable conversations and for sharing his preliminary calculations, as well as Paul Julienne, Chris Greene, Rudolf Grimm, and Christophe Salomon for useful discussions. We acknowledge support from the NSF-MRSEC program, NSF Grant No. PHY-1206095, and Army Research Office Multidisciplinary University Research Initiative (ARO-MURI) Grant No. W911NF-14-1-0003.


%

\renewcommand{\thefigure}{S\arabic{figure}}
\renewcommand{\thetable}{S\arabic{table}}
\renewcommand{\theequation}{S\arabic{equation}}

{\bf Experimental setup and procedure.}~Our experimental apparatus relies on a dual magneto-optical trap (MOT) and dynamic optical dipole traps to produce an ultracold mixture of $^6$Li and $^{133}$Cs. We sequentially load Li and Cs atoms into two independent, spatially separated dipole traps (denoted here as A and B) to avoid light-assisted collisional loss, and adiabatically merge them in a single trap. Both Li and Cs are polarized into their lowest hyperfine states to prevent inelastic two-body collisions. Following the merge process, we compress the trap to maximize the spatial overlap of the two species and prepare the sample at an initial magnetic field of 890~G, where the mixture is stable at a small interspecies scattering length $a_{\rm LiCs} = -21~a_0$, where $a_0$ is the Bohr radius. Then the magnetic field is ramped in $\sim1$~ms to a variable value $B$ near the Feshbach resonance at $B_0$, giving a variable scattering length $a_{\rm LiCs}$.

Trap A is primarily used to confine Li atoms in the initial stage of the experiment and is generated by a pair of 1070~nm laser beams generated from a 200~W Yb fiber laser. The beams propagate parallel to the $x$-axis before the focusing lens, and then are focused at the position of the Li atoms. The beams cross at an angle $\theta = 15^\circ$ at the focus with $1/e^2$ radii of 40~$\mu$m. After loading Li atoms, trap A is spatially displaced up to $25$~mm along the $x$-axis by moving the focusing lens~[27]. This step is essential to avoid light-assisted interspecies collisional loss.

Trap B initially captures laser-cooled Cs atoms, and eventually confines both atomic species. The trap is formed at the intersection of two orthogonally propagating 1064~nm laser beams, whose intensities are independently controlled. The beam propagating along the $z$-axis has a $1/e^2$ radius of 400 $\mu$m, while the beam propagating along the $x$-axis has an elliptical cross section with $1/e^2$ radii of $w_z = 200\;\mu $m and $ w_y = 60\;\mu$m. The elliptical beam is spatially modulated at 1.2 MHz along the $y$-axis (gravity direction) to dynamically increase the vertical size of trap B by about a factor of 3.

\begin{figure*}[t]
\begin{center}
\includegraphics[width=5.5in]{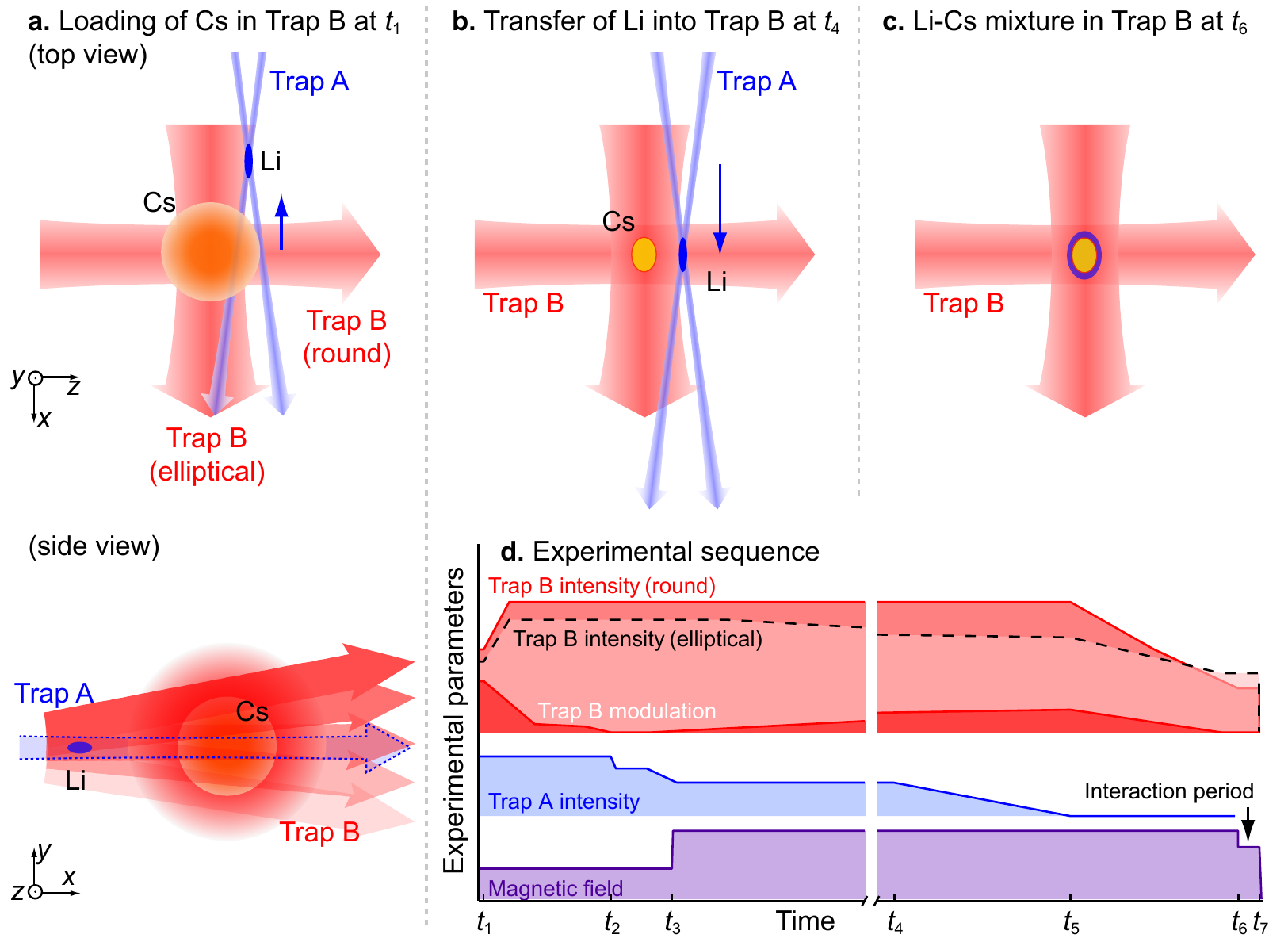}\\
\end{center}
\vspace{-10pt}
\caption{{\bf Schematic of the Li-Cs merge procedure}. The experimental sequence and the meaning of the time points $t_i$ are described in the text. {\bf a.} Schematic of Cs loading in trap B at maximum modulation (side view) while trap A is displaced along $x$. {\bf b.} Location of trap A prior to the transfer of Li atoms from trap A to trap B. {\bf c.} Schematic of the combined mixture in trap B. {\bf d.} Experimental sequence.}
\label{fig:FigureS1}
\end{figure*}

Following the loading and evaporation of Li in trap A, we displace trap A and begin the loading and optical cooling of Cs. After a series of Raman sideband cooling stages Cs atoms are loaded into trap B. We denote the time following the last sideband cooling stage as $t_1=0$~ms. At this point the modulation of the elliptical beam is maximum (Fig.~\ref{fig:FigureS1}(a)) and a magnetic field gradient is applied to levitate only the lowest hyperfine state of Cs, resulting in a high Cs spin purity. Subsequently we compress Cs atoms by reducing the modulation of the elliptical beam and increasing the intensity of trap B. Simultaneously we ramp the magnetic field gradient to zero, until a tight optical trap is achieved at $t_2=2515$~ms. The tighter trap B improves evaporation efficiency, and does not require magnetic levitation. This allows us to relax the Li in trap A from $t_2$ to $t_3=3705$~ms. At $t_3$ we also jump the magnetic field over 25~ms to 890~G where Cs evaporation can be most efficient. From $t_3$ to $t_4 = 10754$~ms we evaporate Cs by increasing the modulation of the elliptical beam and decreasing its intensity. During this time we move trap A into the round beam of trap B (Fig.~\ref{fig:FigureS1}(b)). From $t_4$ to $t_5=11802$~ms we release the Li atoms from trap A into trap B by ramping the intensity of trap A to zero (Fig.~\ref{fig:FigureS1}(c)).  The relative position of the traps before the release of Li atoms is offset vertically by $\sim 100$~$\mu$m and horizontally by $\sim 200$~$\mu$m; these coordinates are optimized experimentally to achieve suitable conditions for the atom-number loss measurements. If trap A is positioned too close to trap B, Cs atoms will leak and cause severe Li loss. On the other hand, if trap A is too far from trap B, the Li transfer efficiency into trap B will be low. Between $t_5$ and $t_6=12711$~ms, trap B intensity is decreased while the modulation is ramped to zero so as to evaporate and compress into a trap where Li and Cs overlap. We also fire a short burst of light resonant with the second-to-lowest Li hyperfine state to remove these unwanted atoms while preserving the Li atoms in the lowest hyperfine state. At $t_6$, the magnetic field is jumped to a variable value, and finally the atoms are imaged at $t_7=12801$~ms. During the interaction period there is a minimum trap depth at which the species can be made to overlap, limiting Cs temperature to about 200~nK. The relevant experimental parameters are depicted in Fig.~\ref{fig:FigureS1}(d). At the final trap where both Li and Cs number are recorded, the trapping frequencies are $(\omega_x,\omega_y,\omega_z)/2\pi=(10,75,20)$~Hz for Cs and $(60,220,120)$~Hz for Li.

\begin{figure*}[t]
\begin{center}
\includegraphics[width=3.5in]{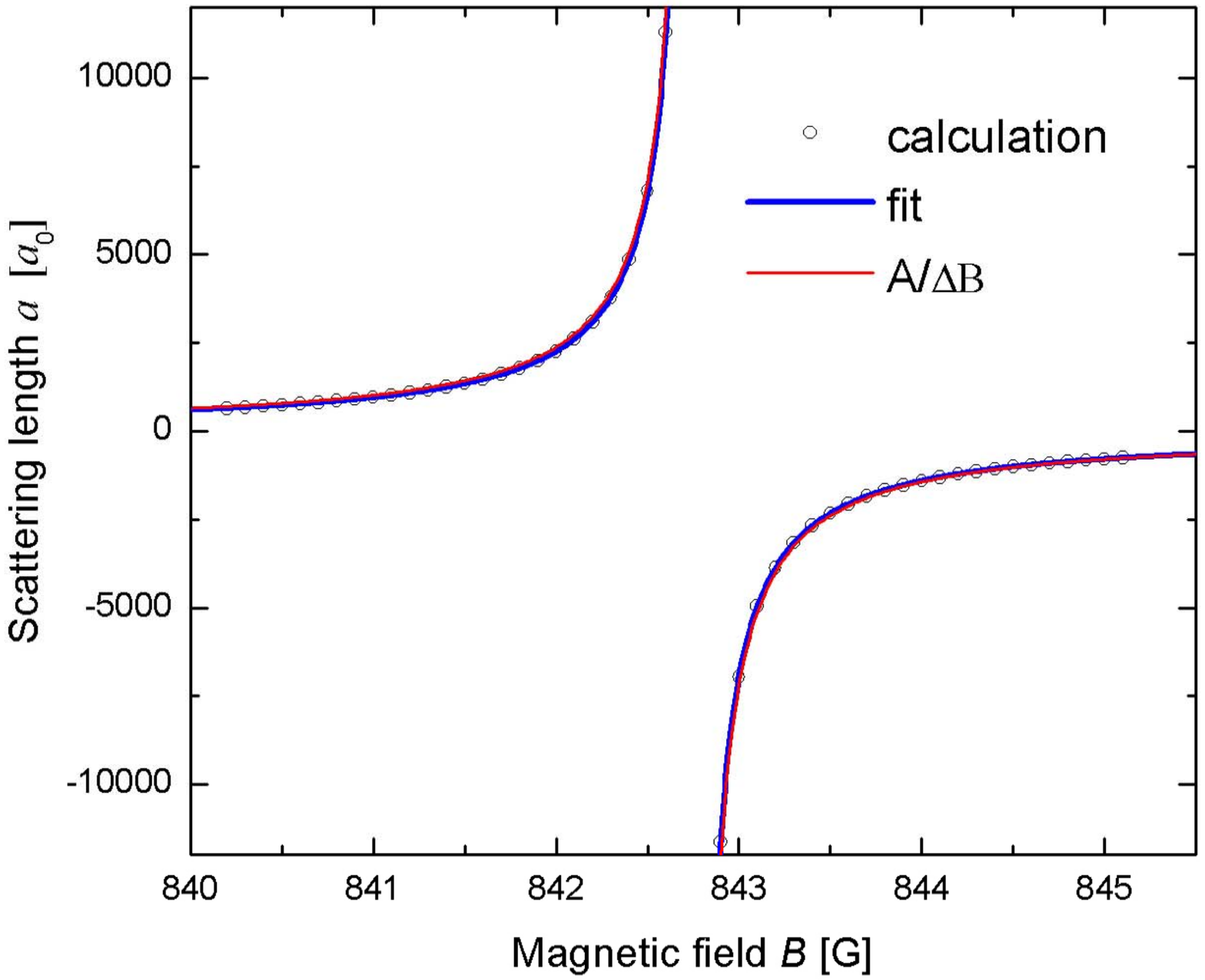}\\
\end{center}
\vspace{-10pt}
\caption{{\bf Scattering length near Feshbach resonance.} Based on the updated value of the Feshbach resonance $B_0=842.75~$G, scattering lengths obtained from a multi-channel calculation~[35] (open circles) are shown, and compared to an empirical fit (Eq.~S1, blue line), and the threshold behavior $a=A/(B-B_0)$, where $A=-1793 a_0 {\rm G}$. In the range of our experiment, the empirical fit is accurate to 99.9$~\%$, and the threshold law is good to better than 95$\%$.}
\label{fig:FigureS2}
\end{figure*}

{\bf Magnetic field calibration and scattering length.}~
We use microwaves to drive Cs atoms from the lowest to the highest hyperfine state of the $6\:{^2\rm S}_{1/2}$ manifold, and convert the observed transition frequency (typically 11.3~GHz at 850~G) into a magnetic field value using the Breit-Rabi formula. The average width of the spectroscopic signals is 30~kHz, corresponding to 12~mG in magnetic field. Day-to-day drifts in the magnetic field of up to 30 mG are corrected by regular calibrations to a precision of $<$ 10mG. We take 30~mG as our conservative estimation on the uncertainties of the absolute magnetic field values reported in this work. To evaluate the uncertainties of the relative separations between Feshbach and Efimov resonances $\Delta B_n = B_n - B_0$, we include both the statistical and systematic uncertainties from $B_n$ and $B_0$.

In the magnetic field range discussed in this work, 800$\sim$900~G, the scattering length can be described to an accuracy of $99.9\%$ by
\begin{equation}
 a(B)=-29.1 a_0 (1+\frac{61.60~\rm G}{B-842.75~\rm G})(1+\frac{2.00~\rm G}{B-892.91~\rm G}),
\end{equation}
where the two parentheses on the right hand side contain the contributions from a 61.60~G-wide resonance around which our experiments were performed, and a narrow one with 2.00~G width. The above expression was obtained from a multichannel molecular model for both Li ad Cs atoms in the lowest internal state~[35], based on the revised Feshbach resonance position $B_0=842.75$~G (see Fig.~\ref{fig:FigureS2}).~We used this expression to calculate the scattering lengths reported in this work.

The three Efimov resonances reported in this work occur within $1/10$ of the resonance width; in this regime, the scattering length can be excellently approximated by $a=A/(B-B_0)$, see Fig.~\ref{fig:FigureS2}. Remarkably, this means that the Efimov geometric scaling law can be recast in terms of the relative magnetic field separations as
\begin{equation}
\Delta B_n\approx \frac{\Delta B_{n-1}}{\lambda},
\end{equation}
where $\lambda$ is the scaling factor. 

{\bf Extracting the three-body recombination coefficient $K_3$ from atom number evolution.}~After each experimental cycle, we perform absorption imaging on both atomic species. Li images are taken \emph{in situ} at a magnetic field of $853$~G, while Cs images are recorded after a 15~ms time of flight (TOF) near zero magnetic field, from which we measure the atom number and temperature.~Modeling the atom number evolution in a Li-Cs mixture needs to take into account several experimental complications. Firstly, the Li and Cs clouds only partially overlap because of the large differences in their masses, magnetic moments, overall trapping potentials and consequently their different gravitational sags in the trap. The typical separation between the two atomic samples is 16~$\mu$m and prevents our experiments from reaching temperatures below 200~nK. (For $T<200$~nK the Li and Cs clouds completely separate.) Secondly, Cs-Cs-Cs recombination plays a crucial role in the evolution of the Cs number, which also  influences indirectly the Li number. Finally, near the Li-Cs Feshbach resonance and Li-Cs-Cs Efimov resonances, fast two-, three-, and higher-body collisions can drive the sample out of thermal equilibrium, which not only can influence the overlap of the clouds, but also can introduce temperature evolution. These effects can significantly contribute to systematic uncertainties in the extracted three-body recombination coefficient $K_3$.

\begin{figure*}[t]
\begin{center}
\includegraphics[width=6in]{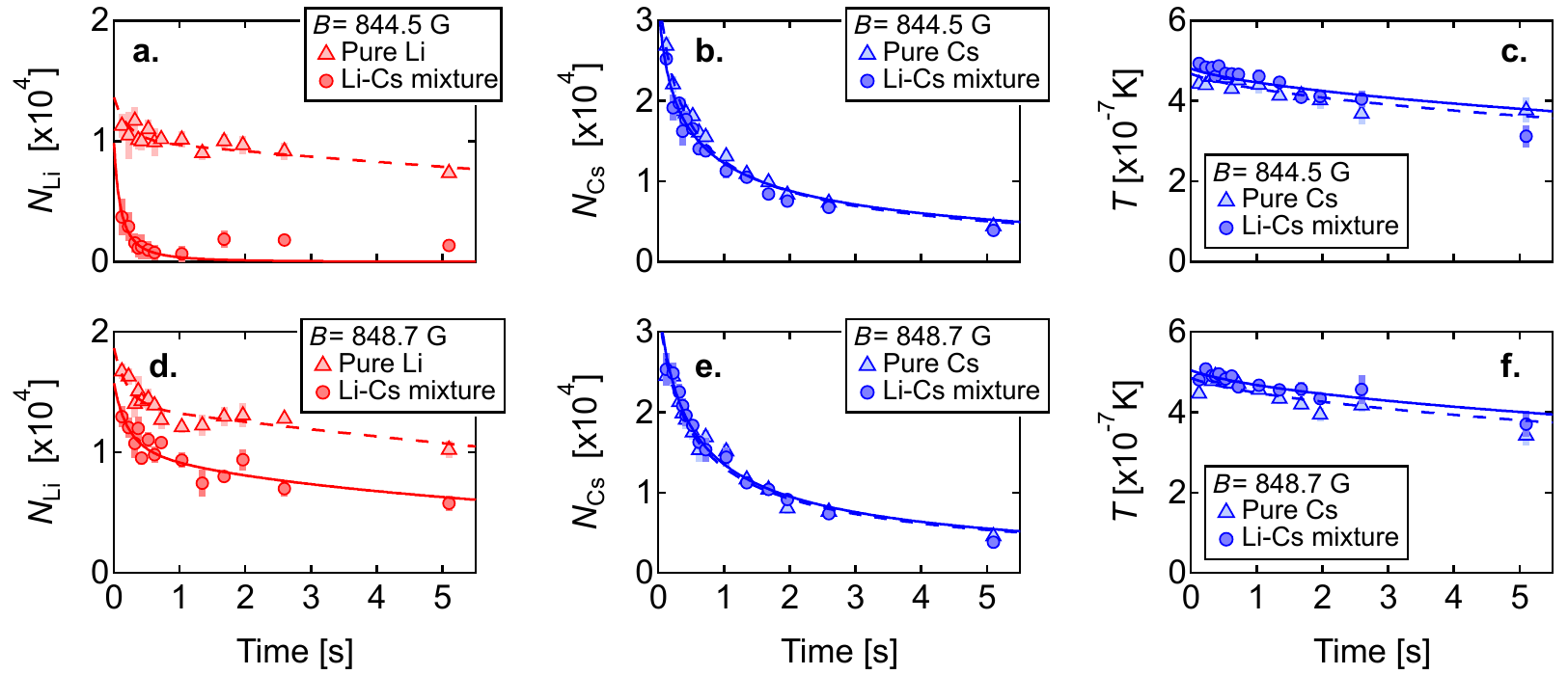}\\
\end{center}
\vspace{-10pt}
\caption{{\bf Temperature and atom number evolution}. Atom number and temperature are measured after different hold times at two different magnetic fields, for mixed and single-species samples. {\bf a.} Li number, {\bf b.} Cs number and {\bf c.} Cs temperature at 844.5 G (near the second Efimov resonance, where the loss rate is high) {\bf d.} Li number, {\bf e.} Cs number and {\bf f.} Cs temperature at 848.7 G (near the first Efimov resonance, where the loss rate is lower).  The continuous and dashed curves represent fitting curves (described in the text) to the mixture data and the single-species data, respectively.}
\label{fig:FigureS3}
\end{figure*}

Despite the aforementioned concerns, we have developed an effective model capable of describing the measured particle number evolutions assuming thermal equilibrium and that loss is dominated by the three-body recombination (Fig.~\ref{fig:FigureS3}). In our model, the atom number $N$ and temperature $T$ are governed by the following rate equations:
\begin{eqnarray}
\frac{{\rm d}N_{\rm{Li}}}{{\rm d}t} &=& -K_3G(T)N_{\rm{Cs}}^2N_{\rm{Li}}-A N_{\rm{Li}}-B \exp(-C t)N_{\rm{Li}},\\
\frac{{\rm d}N_{\rm{Cs}}}{{\rm d}t} &=& -2 \alpha K_3 G(T) N_{\rm{Cs}}^2N_{\rm{Li}}-D N_{\rm{Cs}}^3T^{-3},\\
\frac{{\rm d}T}{{\rm d}t} &=& -F N_{\rm{Cs}}T^{-1},
\end{eqnarray}
where $G(T)$ is a temperature dependent overlap function which we calculate independently, $K_3$ is the desired Li-Cs-Cs loss coefficient, $\alpha$ is the scaling factor to account for atom number calibration error, and $A$, $B$, $C$, $D$, and $F$ are loss parameters. Here $A$ represents one-body losses due to background collisions, while $B$ and $C$ characterize the loss from the Li trap following the final trap adjustment and magnetic field ramping, which perturbs the sample and leads to trap loss during the equilibration process.  Collisions between cold $^{6}$Li atoms are forbidden due to the Pauli exclusion principle. $D$ describes three-body Cs loss, and $F$ describes the decrease of temperature due to Cs evaporation. No evaporative number loss term is added as the number loss of evaporation is negligible compared to three-body loss. Furthermore, since Cs three-body loss also dominates over any non-adiabatic trap ramping or one-body effects only the former is included. The overlap $G(T)$ is given by
\begin{equation}
G(T)= \frac{1}{N_{\rm{Cs}}^2N_{\rm{Li}}}\int n_{\rm{Cs}}(\mathbf{x},T)^2 n_{\rm{Li}}(\mathbf{x},T)\,{\rm d}^3\mathbf{x} ,
\end{equation}
where $n$ represents density and the integration runs over the whole trap. The sample density distributions are assumed to be in equilibrium at temperature $T$ for simplicity. The density profiles are derived from the measured trapping frequencies and gravitational sags.

The constants $A$, $B$, $C$, $D$, and $F$ are obtained from fits to decay profiles from pure Li and Cs samples in isolation. The data at all fields are well described with the same parameters. Subsequently, $K_3$ is determined from fits to decay profiles of mixed samples. We obtain $\alpha = 0.26$, which differs from one due to imperfect particle number calibration, and because the Li-related loss of Cs is difficult to separate from the high background rate of Cs loss. This has little effect on $K_3$ however, as $K_3$ is determined primarily from the Li loss. With this model, we can determine the expected number of Li atoms remaining after a fixed hold time as a function of $K_3$, and the initial atom numbers and temperature (Fig.~\ref{fig:FigureS3}). For each point of a number loss trace (Fig.~3), we interpolate $N_{\rm Li}$ to find its predicted $K_3$ value. We estimate that the derived recombination coefficients $K_3$ can suffer from an overall scaling uncertainty on the order of 10. In the main manuscript, Fig.~4 shows our calculated dimensionless three-body loss coefficient $K_3/K_0$ as a function of magnetic field, where $K_0=10^{-25} {\rm cm}^6/{\rm s}$ is the loss coefficient we obtain at $\sim$852~G.

\clearpage
{\bf Determination of the Efimov and Feshbach resonance positions.}~In order to obtain the necessary statistics to precisely identify the Efimov resonances, we averaged many scans taken over several days. As a result, drifts in experimental parameters can have significant effects on atom number. To remove these drifts, we rescale our data so that it averages to one over a fixed magnetic field range. Small drifts in the magnetic field (on the order of 10 mG over 6 to 8 hours) mean that nominally identical scans taken at different times will include points at different values of the true magnetic field. In order to average these data together, we employ an interpolation scheme. After rescaling and interpolating, we remove the outliers of the data points and average our data. The removal of outliers shifts the resonance position by no more than 5 mG.

\begin{figure*}[t]
\begin{center}
\includegraphics[width=6in]{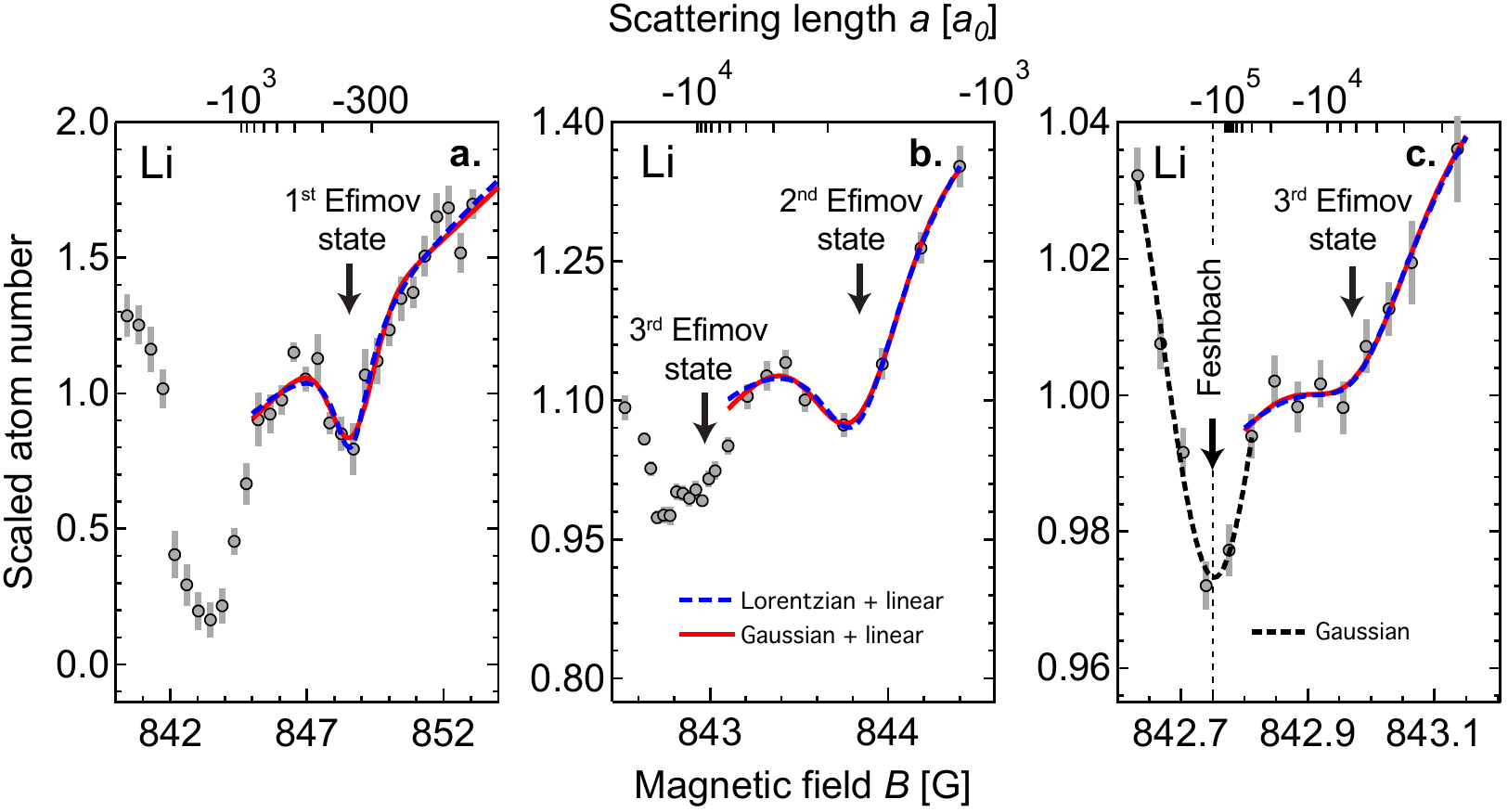}\\
\end{center}
\vspace{-10pt}
\caption{{\bf Determination of the Efimov and Feshbach resonance positions.}~Based on the atom number data in Fig.~3, the locations of Efimov resonances are determined from a superposition of linear and Gaussian fits (continuous lines) and from a superposition of  linear and Lorentzian fits (dashed lines). A Gaussian fit to the Feshbach resonance is also shown in panel c (dotted line).}
\label{fig:FigureS4}
\end{figure*}

Since analytic forms of the Efimov resonance lineshape at finite temperature are unavailable, we used empirical fits to determine the positions of Efimov resonances. Fit functions based on a Gaussian or a Lorentzian plus a linear background were adopted, see Fig.~\ref{fig:FigureS4}. ~We also determined the peak positions from fitting the derived recombination coefficient $K_3$ to a Gaussian with a linear background. We also determine the decay parameter $\eta$ of the first Efimov resonance, where the finite temperature effect is minimal. Using the analytic function in Ref.~[37]  to fit the recombination coefficient $K_3$, we obtain $\eta$ = 0.26(6).

In addition to different fit functions, we also repeated those fits with peripheral data points excluded to determine their influence on resonance positions and uncertainties. All fits gave consistent results. We defined the final resonance position for each resonance as the mean of all fits. Furthermore,  the final uncertainty for each resonance was determined by the square root of the quadrature sum of the mean statistical uncertainty and the standard deviation of all fitted resonance positions. Our final values and uncertainties are summarized in Table~\ref{tab1}.

\begin{table*}[t]
\caption{Determination of Efimov resonance positions using different methods}
\center
\begin{tabular}{c |c c | c | c | c}
  \hline
  Efimov & \multicolumn{2}{c|}{Atom loss}&\multicolumn{1}{c|}{$K_3 $}&& \\
  \;\;resonance\;\; & \;\;Gaussian\;\;  & \;\;Lorentzian\;\; & \;\;Gaussian\;\; & \;\;Final value\;\; & \;\;Scattering length\;\; \\
  \hline
  $B_1$ [G] & 848.60(11)& 848.55(11)& 848.49(12)& 848.55(12) & -323(8) $a_0$\\
  $B_2$ [G] & 843.82(5)	& 843.83(5)& 843.81(3)& 843.82(4) & -1635(62) $a_0$\\
  $B_3$ [G] & 842.97(3)	& 842.97(3)& 842.97(2)& 842.97(3) & -7850(1124) $a_0$\\
  \hline
\end{tabular}\\
\vspace{5pt}
The number in parenthesis represents statistical uncertainty. Systematic uncertainties are 30~mG.
\label{tab1}
\end{table*}

To determine the Feshbach resonance position, we employed two independent methods. The first one was based on the position of the strongest dip in the atom loss spectrum. We assign this feature as the Feshbach resonance position, since it persists in both Li and Cs data at all temperatures (up to 800~nK in this work and few $\mu$K in previous work [27]). It exists even when the Efimov features are significantly weakened above 500~nK. We consider that the strongest dip may be coming from fast evaporative loss due to resonantly-enhanced two-body collisions. Indeed, at temperatures around 250~nK, Li-Cs two-body collisions reach the unitary limit only at scattering lengths $|a|>1/k=11000$~$a_0$ or within 150~mG of the Feshbach resonance. Here $k$ is the Li-Cs relative wavenumber. 

\begin{table*}[t]
\caption{Determination of the Feshbach resonance position using different methods}
\center
\begin{tabular}{c | c c |c | c}
  \hline
  Feshbach& \multicolumn{2}{c|}{Atom loss dip}  & \multicolumn{1}{c|}{$K_3$}& \\
  \;\;resonance\;\;  & \;\;Gaussian\;\; & \;\;Lorentzian\;\;& \;\;Steep slope\;\; & \;\;Final value\;\; \\
  \hline
  $B_0$ [G] & 842.75(1) & 842.75(1) & 842.75(1) & 842.75(1) \\
  \hline
\end{tabular}
\vspace{5pt}
\label{tab2}\\
The number in parenthesis represents statistical uncertainty. Systematic uncertainties are 30~mG.
\end{table*}

\begin{figure*}[b]
\begin{center}
\includegraphics[width=6in]{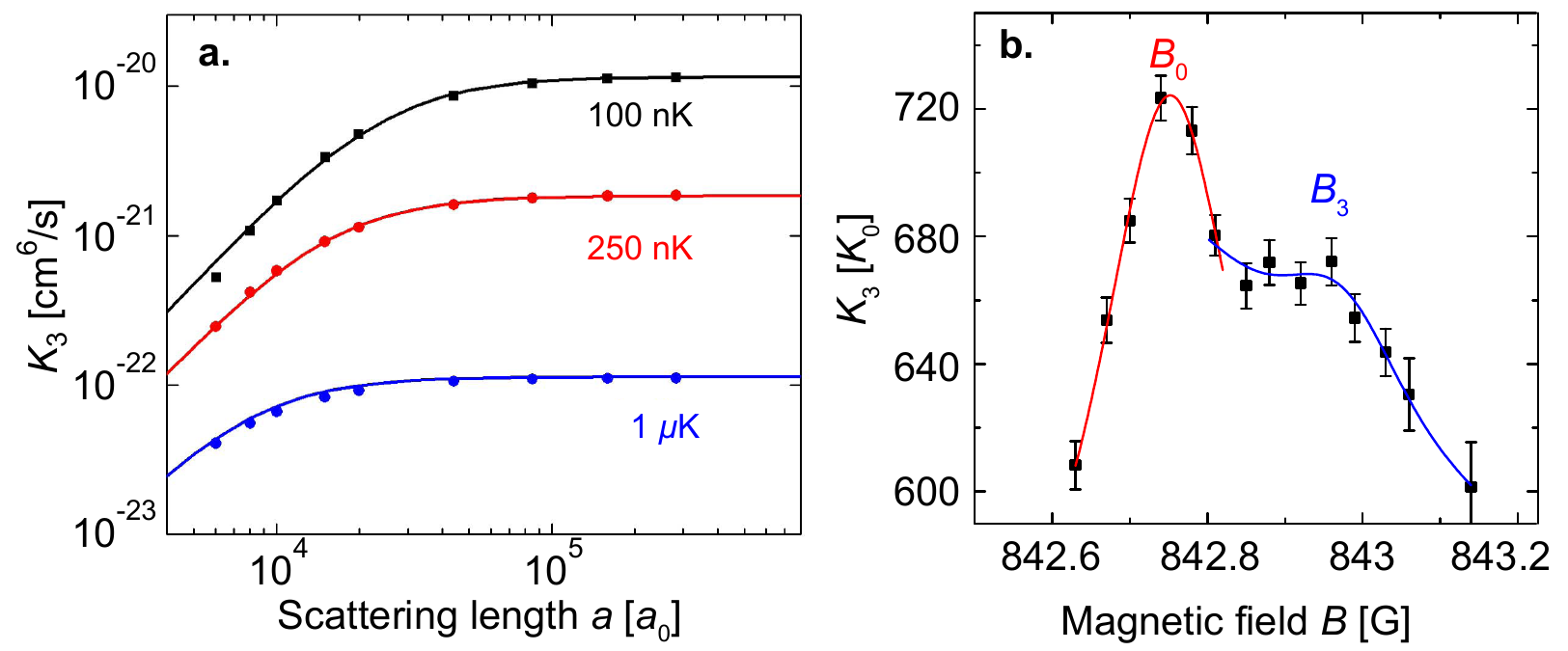}
\end{center}
\vspace{-10pt}
\caption{{\bf Steep rise of the recombination coefficient $K_3$ for $a>0$.} {\bf a.} Numerical calculation for $a>0$ (solid dots) [36] are fitted to an empirical function of $K_3=Aa^2{k^*}^{-2}/(1+{k^*}^2a^2)$ (continuous curves). With fitting parameters $A=1154 \hbar^2/m_{Li}$ and $k^*=(T/$nK$)^{1/2}(2.41\times10^5~a_0)^{-1}$, the fit function well captures the calculations at all three temperatures. {\bf b.} Based on the fit function, including an overall scaling as the third fitting parameter, we fit the data on the steep slope of the extracted recombination rate $K_3$ at $T= 270~{\rm nK}$ and obtain $B_0 = 842.75(1)~{\rm G}$ (red curve). A Gaussian fit to the third Efimov feature is shown for comparison.}
\label{fig:FigureS5}
\end{figure*}

An independent method to determine the Feshbach resonance position is to use the steep rise of the loss on the positive scattering length side to fit the experiment data. This steep slope, clear in both experiment  (Fig.~\ref{fig:FigureS5}) and calculation [36] (Fig.~\ref{fig:FigureS7}), allows us to determine the Feshbach resonance position with high precision and, in principle, only requires data below the Feshbach resonance.

To apply this idea, an empirical function, interpolating the asymptotic behavior $K_3\sim T^{-2}$ on resonance and $K_3\sim a_{\rm LiCs}^2$ in the threshold regime, is found to well fit the calculations at all temperatures (see Fig.~\ref{fig:FigureS5}). Armed with the fit function, we fit the recombination loss data near the Feshbach resonance at $T=270$~nK by taking the Feshbach resonance position $B_0$ and the overall scale of the loss as the only two fitting parameters.  We find that the function captures the data very well for $a>0$, see Fig.~\ref{fig:FigureS5}(b), and the resonance position is determined to be $B_0=842.75(1)_{\rm stat}~$G, see Table~\ref{tab2}. Our result is consistent with the dip position measurement. It, however, deviates from our previous work by $-0.65$~G [27]. We attribute the deviation to the much higher temperatures reported of few $\mu$K in our former work, where the Efimov features were indiscernible, resulting in a single smooth and broad loss profile.

\begin{figure*}[b]
 \begin{center}
 \includegraphics[width=3.5in]{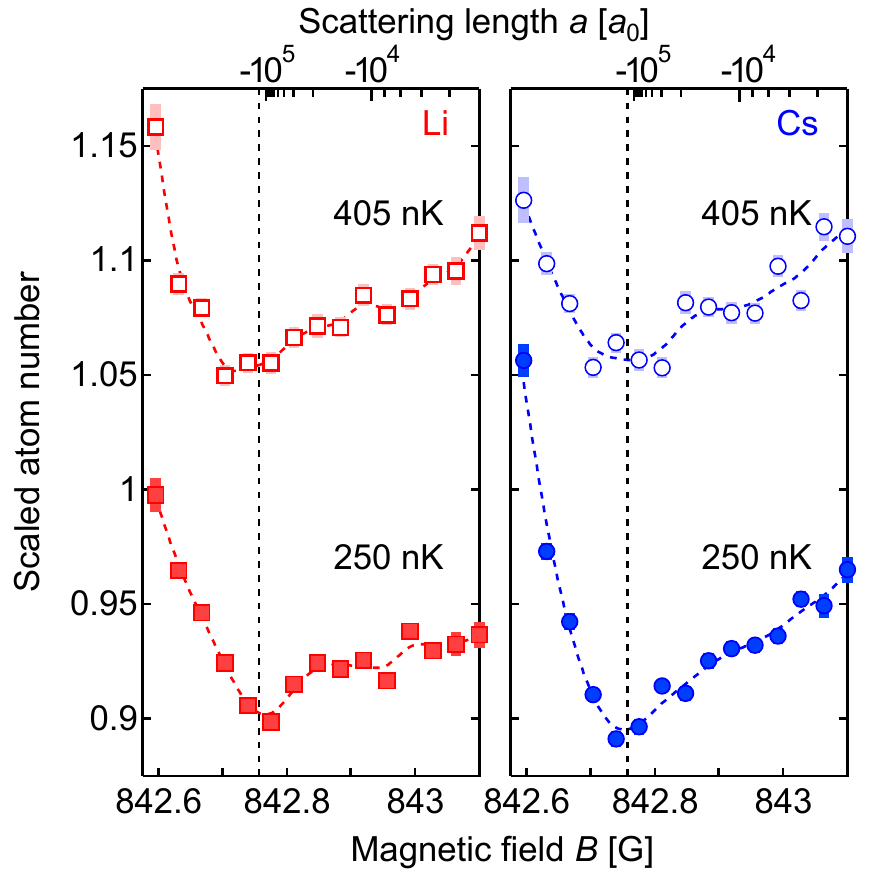}\\
 \end{center}
\vspace{-10pt}
\caption{{\bf Third Efimov resonance at different temperatures}. Scaled Li (squares) and Cs (circles) numbers versus magnetic field at mean temperature $T= 405$~nK (top) and 250~nK (bottom). The mean atom numbers are $N_{\rm Li} = 1.4\times 10^4$ and $N_{\rm Cs} = 1.3\times 10^4$ (top), and ${N_{\rm Li} = 1.3\times 10^4}$ and $N_{\rm Cs} = 7.0\times 10^3$ (bottom). The scaled atom numbers come from the average of 209 (top) and 328 (bottom) traces normalized to their respective means. The typical hold time is {115 ms}. The vertical dashed lines indicate the Feshbach resonance. The traces are offset vertically by 0.15 for clarity, and their interpolation (dashed curves) serve as guides to the eye.}
\label{fig:FigureS6}
\end{figure*}

{\bf Finite temperature and finite trap volume effects.}~Signatures of higher order Efimov resonances suffer from finite temperature effects when the thermal de Broglie wavelength $\lambda_{\rm dB}$ of the atoms is small compared to the scattering length at which the resonance occurs $a_-^{(n)}$. At the lowest temperature reported here (250~nK) $\lambda_{\rm dB}^{\rm Cs}=5700~a_0$ and $\lambda_{\rm dB}^{\rm Li}=27000~a_0$, which are both much larger than the second Efimov resonance position $|a_-^{(2)}|=1635~a_0$, but comparable with that of the third resonance $|a_-^{(3)}|=7850~a_0$. This is consistent with our capability to see only three resonances. In addition, finite temperature can lead to a shift of the resonance position.

Experimentally we investigated the temperature dependence of the third Efimov resonance by performing experiments with temperatures $T=190\sim500$~nK. We separated the data into two groups with mean temperatures $T$ = 250 and 405~nK, see Fig.~\ref{fig:FigureS6}. Because of the limited data quality, we cannot conclude whether there is a temperature shift. The data, however, confirms that the third Efimov resonance feature is weaker at high temperature, consistent with Fig.~3. 

\begin{figure*}[b]  
\begin{center}
\includegraphics[width=3.5in]{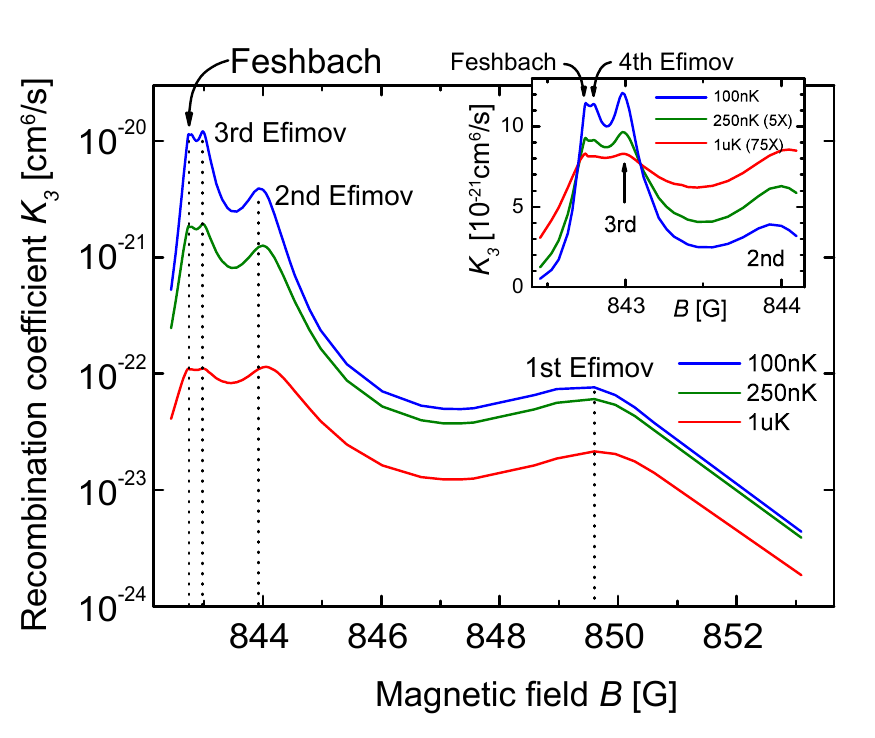}\\
\end{center}
\vspace{-10pt}
\caption{{\bf Theoretical calculation of the Li-Cs-Cs three-body recombination coefficient}. The calculation, conducted by Y. Wang~[36], is based on the revised location of the Feshbach resonance reported in this work and other resonances reported previously~[27]. The enhancements of the recombination coefficient indicate Efimov resonance positions (dashed lines) and correspond to the dips in the atom number shown in experiment (Fig.~3).As temperature increases, the calculation suggests a weaker resonance features and shifts of the resonance positions towards higher magnetic fields. The inset shows the same data in linear scale near the Feshbach resonance. A signature of the fourth Efimov state is predicted at low temperatures and is located at $\sim$40~mG above the Feshbach resonance.}
\label{fig:FigureS7}
\end{figure*}

Complementary to our experiment, an independent three-body calculation conducted by Y. Wang shows the expected recombination loss coefficient in the temperature range of interest [36] (Fig.~\ref{fig:FigureS7}). The calculation shows signatures of Efimov resonances at various temperatures, and confirms the suppression of the resonance features at higher temperatures. Based on a Gaussian plus linear fit to the calculation, we find that the resonance position shifts in the range of 100-1000~nK are 210~mG, 95~mG and 7~mG for the first, second and third resonances, respectively. Those shifts are comparable to the experimental uncertainties of the first and the second resonance; however, they do not lead to significant corrections in the scaling constants we derived. We estimate that the finite temperature effects in our experiment can offset the scaling ratios $\lambda_{21}$ and $\lambda_{32}$ by -1$\%$ and -3$\%$, respectively, which are both small compared to our quoted uncertainties of 4$\%$ and 15$\%$.

The calculation also supports the identification of the Feshbach resonance position based on the leftmost resonant loss feature, see Fig.~\ref{fig:FigureS7} inset. At experimental temperatures, the fourth Efimov state does not form a clear resonance feature, consistent with our observation, but can lead to an asymmetric lineshape. Since we determine the same Feshbach resonance position using data taken on the positive scattering length side (Fig.~\ref{fig:FigureS5}(b)), as by fitting the loss feature directly, we conclude that our procedure to determine the Feshbach resonance position only suffers weakly from the asymmetric line shape with systematic uncertainty of 10~mG or less.

The trapping potential can also affect the position of higher Efimov states if the Efimov binding energy approaches the trap vibrational frequency. In this case Efimov trimers bind not relative to the scattering continuum, but relative to the discrete set of bound states determined by the trap. The binding energy scale for heteronuclear Efimov trimers is $\hbar^2/(\mu a_{\rm LiCs}^2)$, with $\mu = [m_{\rm Li}m_{\rm Cs}^2/(m_{\rm Li}+2m_{\rm Cs})]^{1/2}$ the three body reduced mass~[36]. In the case of the third Efimov state we observed here, this energy scale is $h\times$3~kHz, much greater than the largest trap vibrational energy of Cs ($h\times$75~Hz) or that of a bound LiCs molecule ($h\times$90~Hz). [For LiCs molecules, we assume the polarizability of the dimer is the sum of the atoms and the trap frequency is given by $(m_{\rm Li}+m_{\rm Cs})\omega_{\rm LiCs}^2 = m_{\rm Li}\omega_{\rm Li}^2+m_{\rm Cs}\omega_{\rm Cs}^2$.] We therefore conclude that the trap effect is negligible.


\begin{thebibliography}{34}
\bibitem[]{note1}\BibitemOpen{In our manuscript, Efimov resonances represent exclusively those three-body recombination loss resonances with negative scattering lengths.}\BibitemShut {NoStop}%
\end{thebibliography}
\end{document}